# Patterns of Social Vulnerability – An Interactive Dashboard to Explore Risks to Public Health on the US County Level

Darius Coelho*  
Akai Kaeru LLC

Nikita Gupta†  
Stony Brook University

Eric Papenhausen‡  
Akai Kaeru LLC

Klaus Mueller§  
Akai Kaeru LLC  
Stony Brook University

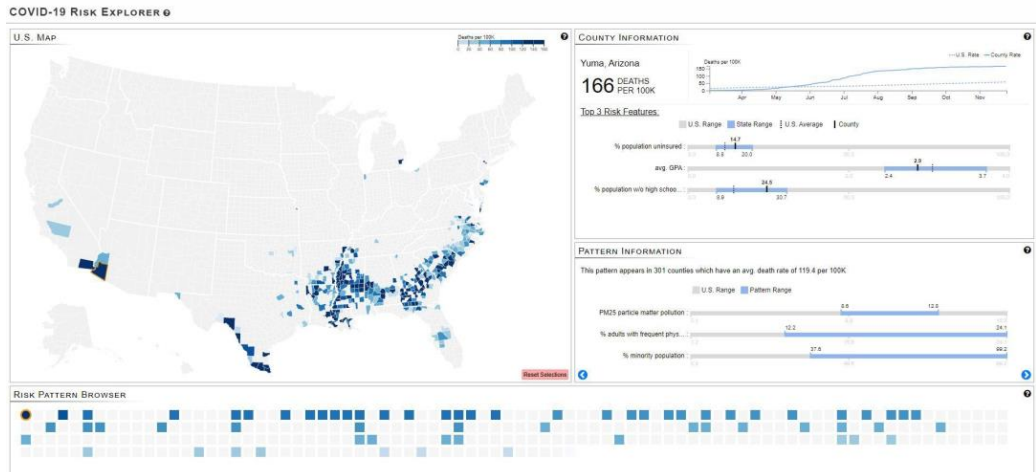

Figure 1: The COVID-19 Risk Explorer dashboard for exploring risk patterns that drive COVID-19 death rates in the US. On the the top left is an interactive geomap of the US counties shaded by their COVID-19 death rate. A county can be selected by clicking the mouse on its map location. Here, the user selected Yuma, AZ, marked by an orange outline. The bottom panel has the risk pattern browser which represents all the risk patterns as tiles, shaded by their COVID-19 death rate. Only the patterns relevant to the selected county (Yuma, AZ) are shaded. Users can select a risk pattern by clicking the mouse on a pattern tile. Here the user selected the first pattern, marked by an orange outline. All counties that share this pattern are then shaded on the map. The top right panel shows information about the selected county while the panel below it offers information about the selected pattern.

## ABSTRACT

Social vulnerability is the susceptibility of a community to be adversely impacted by natural hazards and public health emergencies, such as drought, earthquakes, flooding, virus outbreaks, and the like. Climate change is at the root of many recent natural hazards while the COVID-19 pandemic is still an active threat. Social vulnerability also refers to resilience, or the ability to recover from such adverse events. To gauge the many aspects of social vulnerability the US Center of Disease Control (CDC) has subdivided social vulnerabilities into distinct themes, such as socioeconomic status, household composition, and others. Knowing a community's social vulnerabilities can help policymakers and responders to recognize risks to community health, prepare for possible hazards, or recover from disasters. In this paper we study social vulnerabilities on the US county level and present research that suggests that there are certain combinations, or patterns, of social vulnerability indicators into which US counties can be grouped. We then present an interactive dashboard that allows analysts to explore these patterns in various ways. We demonstrate our methodology using COVID-19 death rate as the hazard and show that the patterns we identified have high predictive capabilities of the pandemic's local impact.

*e-mail: dcoelho@akaikaeru.com  
†e-mail: nikigupta@cs.stonybrook.edu  
‡e-mail: epapenha@akaikaeru.com  
§e-mail: mueller@cs.stonybrook.edu

## 1 INTRODUCTION

Social vulnerability gauges the socioeconomic and demographic factors that affect the resilience of a community to external disastrous events affecting human health, called *stresses*. These stresses can range from natural or human-caused disasters to disease outbreaks. A socially vulnerable community is less likely to recover and more likely to perish as a result of these stresses, and by reducing social vulnerability, one can lower both human suffering and economic losses. Knowing the specific social vulnerabilities for a given community can help emergency response planners and public health officials to quickly respond when a specific disaster strikes and to build long-term residence for it to weaken its potential impact.

The concept of social vulnerability has been studied world-wide and various measures have been established. In the US, the Agency for Toxic Substances and Disease Registry (ATSDR) and the Center of Disease Control (CDC) have used US Census data to determine a Social Vulnerability Index (SVI) for every US census tract. Census tracts are subdivisions of US counties and there are 3,006 counties in the US. The CDC/ATSDR SVI ranks each tract on 15 social factors which can be grouped into into four related themes:

- Socioeconomic status: below poverty, unemployed, income, no high school diploma
- Household composition & disability: aged 65 or older, aged 17 or younger, older than age 5 with a disability, single-parent households
- Minority & language: minority, speak English "less than well"
- Housing type & transportation: multi-unit structures, mobile homes, crowding, no vehicle, group quarters

Using a dataset from Kaggle [11] we have expanded this list to a carefully curated set of 241 factors that refines these measures to demographic features, such as race and gender. and adds further health and social risks, such as smoking and drug use habits, teenage pregnancies, sleep deprivedness, housing debt, vaccination rate, and many others. The Kaggle dataset is composed of data collected from 200 publicly available COVID-19 related datasets, using sources like Johns Hopkins, the WHO, the World Bank, the New York Times, and many others. We found several redundancies in this dataset and carefully trimmed it to the set of 241 factors which are the basis of the methodology and study reported in this paper.

Visualizing the social vulnerabilities as a choropleth map can be helpful to planners, responders, and policymakers to compare the social vulnerability index across locations and so better understand which communities are most susceptible to natural disasters and disease outbreaks. Most prominent is the CDC's own SVI Interactive Map [10] which colors counties by their overall SVI score - the score is a value between 0 (lowest vulnerability) to 1 (highest vulnerability); clicking the mouse on a specific county pops up a scorecard that lists the score for each theme as a bar chart. The US Federal Emergency Management Agency (FEMA) [6] provides a series of choropleth maps that aside from maps for social vulnerability and community resilience also allows users to produce maps for the risks and expected annualized loss for a variety of natural hazards, such as flooding, lightning, tornadoes, and many others. A service called County Health Rankings [3] allows users to produce detailed factor-based comparative reports with visualizations for US states at the granularity of counties. There are also several other institutions, such as NASA, that produce factor-based SVI choropleth maps.

To the best of our knowledge there is no work thus far that allows users to explore the multivariate nature of social vulnerabilities more directly, in the form of patterns of multiple interacting factors. The currently available maps all require users to switch from map to map or perform side by side comparison to assess these correlations. Our pattern analysis groups counties, which are not necessarily geographically related, in terms of vulnerability factors that cooperate in making these counties more (or less) susceptible to a certain target hazard. Enabling analysts to explore these patterns and the locations of the affected counties on a single choropleth map allows for a deeper and more efficient analysis. This paper describes the outcomes of our research geared toward achieving such a map.

Our paper is organized as follows. Section 2 presents related work. Section 3 describes our pattern analysis. Section 4 explains our dashboard that allows users to explore these patterns in the context of the US counties. Section 5 offers a conclusion. The appendix presents a case study.

## 2 RELATED WORK

Unlike previous large-scale global health crises the COVID-19 pandemic arrived in an era with an ubiquity of machine learning tools and a widely developed information technology infrastructure. The urgency of COVID-19 did not only boost efforts in biotech to develop vaccines and medical treatments at rapid speeds, but it also invigorated research in the predictive modeling of the spread of the virus and in the development of visual information portals to inform policy makers and the general public on death tolls, infections, and the like. As such the COVID-19 health emergency brought about much of the most recent related work on visual tools for public health monitoring and risk assessment.

### 2.1 Modeling and Prediction

Many predictive modeling approaches (such as [13], [24]) are based on the mechanistic Susceptible–Exposed–Infectious–Recovered (SEIR) compartment simulation model that, at the process level, mimics the way COVID-19 spreads. Mechanistic models are attractive since they allow one to simulate the effects of different mitigation measures, such as quarantining, social distancing, school closings, and so on. However, the model's many parameters require accurate estimates of the population in each compartment and their transition rates which can be challenging due to the uncertainties involved.

Popular are also statistical forecasting models such as that by U Washington's Institute for Health Metric and Education (IHME) [1]. It uses a mixed effects nonlinear regression model to fit a curve to data from world-wide geographical locations to create projections of infections, death rates, and health resource demands at the local level. Other approaches use more conventional statistical models, such as correlation and linear regression to understand the influence of certain socio-economic factors, such as county-level health variables, urban density, poverty, commuting, and so on, while controlling for other effects, such as race [18]. Typically these results are obtained via standard step-wise modeling approaches that are not overly scalable in the number of factors and local regions, making the discovery of significant statistical relationships rather tedious.

Recognizing that models often produce a wide range of predictive forecasts, ensemble methods have recently become popular (see for e.g. [17]). Ensemble methods combine different individual models together and weigh their outcomes into a unified forecast. This can make predictions more robust and add stability to the process.

### 2.2 Visualization

A primary source of information has been the Coronavirus Resource Center at Johns Hopkins University [4]. They constructed dashboards for the US and for the entire world that each showed the respective geographic maps overlaid with visual representations of the numbers of people tested positive alongside various test and death statics, leader boards, and temporal growth curve ensembles that compare regions at various scales in terms of the increase of test cases and deaths. Other dashboards and browser-based interactive visualization of COVID-19 related data have been made available by the companies Tableau [7], TIBCO [2], the open source project Nextstrain [9], newspapers like the New York Times, and others. These dashboards and visualizations illuminate specific aspects related to the outbreak, such as race, hospital overcrowding, test statistics by state, mask compliance by county, unemployment rates and claims, economic inequality, pathogen evolution, and more.

The IHME COVID-19 forcasting model has also become quite popular due to its simple yet effective interactive visual dashboard tools [5]. However, IHME is also known for its interactive 'US Health Map — Viz Hub' [8]. This visual interface provides a menu that allows users to select among four outcome or risk variables, i.e. life expectancy, mortality rate, mortality risk, and others. Users can then choose one of many diseases and health determinants and display the outcome or risk on a choropleth map for states or counties. However, similar to other maps, their map also can only display one quantity at a time and so enables comparisons only on a factor-wise basis. While juxtaposition [14] of several maps can facilitate comparisons, it remains difficult to recognize general correlations or groupings among the variables. Our geo-display enables it by pairing pattern analysis with a set of information displays.

A recent visual interface is EnsembleVis [21] which is a web-based geomap interface to view and compare model forecasts with the ground truth on the US county level. It allows users to navigate the ensemble models and so gain a better understanding of the ranges and uncertainties. We also display our data on the county level but our focus is mainly to explain *why* certain health risks occur, that is, what are the social vulnerabilities that expose certain communities to greater risk. While our method learns these risks from communities that have already been exposed, there are others that fit this patterns as well but have not suffered the same fate as yet. As such the patterns we identify can also be used to predict or at least alert these communities and associated responders.

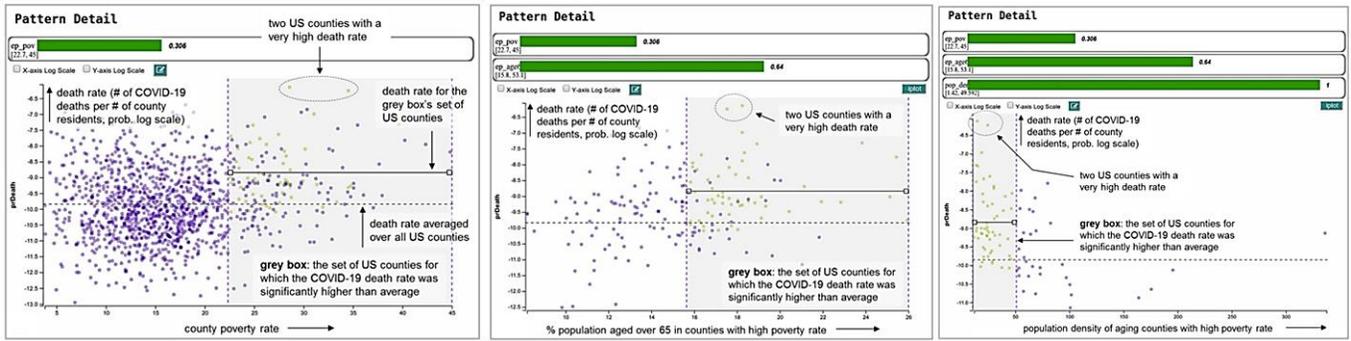

Figure 2: Scatterplot of a 3D pattern found in May 2020. The y-axis (target attribute) is the observed COVID-19 death rate and each point is a US county. Each image shows the 2D projection into the plot's pattern attribute and the target attribute after culling the points into the subspace defined by the pattern attribute of the plot on its immediate left. The yellow points are inside the plot's pattern subspace and the purple points are outside of it. It can be seen that adding the third attribute is sufficient to eliminate all purple points. The culling of points is the reason why there are progressively fewer points in the plots from left to right. The green bars on top show how much each of the subspace attributes (top to bottom, and left to right in the plots: county poverty rate, % population aged over 65, population density) contributes the definition of the pattern.

## 3 OUR APPROACH: PATTERN MINING AND DASHBOARD

In the following we summarize our pattern mining approach and then focus on the dashboard we designed for exploring these patterns.

### 3.1 Our Pattern Mining Approach

Our approach is rooted in pattern analysis, a well-studied area of research in data science and AI [19]. A pattern is a subgroup of data points that share similar characteristics, or features [12]. For example, the data could be a set of counties that have a similar socio-economic make-up. In our example each county has 241 features, e.g., % adults w/excessive drinking habits, % adults in frequent mental distress, unemployment rate, etc. These 241 features then result in a 241-D feature space which is typically fairly sparse.

We have devised a pattern mining engine that automatically searches this sparse feature space for regions occupied with similar counties which all respond in a similar way to a given target variable of interest. Some results of our pattern mining approach are reported in [20]. In that work we specifically focused on COVID-19 and the visualizations consisted of a simple choropleth map which did not offer any capabilities to explore the patterns in terms of their features. We now report on a dedicated dashboard that puts the human in the loop and allows for detailed interrogations.

In our prior work we sought to identify the socio-economic conditions that underlie higher than average COVID-19 death rates, and so our target variable was a county's COVID-19 deaths rate. We note that counties that are considered similar do not need to be geographically connected; they just need to have similar characteristics in terms of their feature values. A unique property of a pattern is that it fits inside a hypercube with well-defined value ranges of the features that describe it. This property and its inherent low dimensionality [22], even when the overall feature space is not, makes them easy to understand and explain. While deep neural networks, random forests, etc. also learn low-D representations, these are not easily described in terms of the native features. Hence, these types of architectures are commonly referred to as black-box AI while pattern analysis is an explainable AI approach.

Concretely, given a dataset with attributes $\{A_1 A_2 ....A_m P\}$ with $P$ being an attribute of interest, such as COVID-19 death rate, the goal of pattern mining is to find a hypercube (or pattern) consisting of constraints of the form $A_i \in [v_l, v_r]$ for $i \in [1...m]$ (for example, age > 45, race = Asian), where the points within the pattern are "interesting." For our purposes, a pattern of counties will be considered interesting if it is associated with a COVID-19 death rate that is higher on average than the US county average. The definition of what constitutes a consistently interesting pattern is based primarily on statistical hypothesis testing. For numerical attributes, we use the Mann-Whitney test [23] to account for the often non-parametric nature of the data, while for a binary target attribute, we use the $\chi^2$ test for independence. Extracting the patterns requires extensive search; we use the FP-growth algorithm [16] which is fairly efficient as it only requires scanning the full dataset twice during the mining.

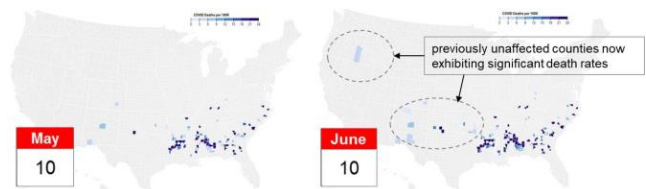

Figure 3: The locations of the counties in the pattern of Fig. 2. colored by COVID-19 death rate. The newly disease-stricken counties in June 2020 (inside the dotted ellipses but also elsewhere) are counties located in the bottom of May 2020's scatterplot in Fig. 2 or not yet visualized there, but correctly predicted to get hit soon.

### 3.2 Our Pattern Mining: A Closer Look

For our prior COVID-19 study we used the 241-D dataset mentioned in the introduction and used COVID-19 death rate in each US county as the target variable. We found 297 2D and 3D patterns in May 2020. Fig. 2 visualizes one of them, a 3D pattern defined by high poverty rate, high percentage of senior citizens, and low population density. These three variables were sufficient to confirm the statistical significance for the elevated death rate average. The figure caption explains the formation of the pattern in greater detail.

Let us have a closer look at the plot on the very right of Fig. 2 which shows the projection of US counties into the pattern's 3rd attribute (yellow points). We notice that most US counties in the pattern have a death rate above the US average (and the pattern's average itself is also above the US average which makes the pattern "interesting"). But we also notice that there are a few US counties that are below the US average bar; they have a death rate below the US average. This can mean that there are other latent (unmeasured) factors that protected the counties from contracting the virus. But it can also mean that these counties were not yet hit by the COVID-19 wave – recall that May 2020 was very early in the pandemic.

Fig. 3 gives more insight into this. It shows two maps where the counties with significant death rates are shaded in blue – deeper blue shades map to higher death rates. Note that we only colored counties that matched the pattern shown in Fig. 2 (they may also match other patterns but we did not consider these for this plot). On the left is the map for May 2020, the month we used to learn our patterns. On the right we see the corresponding map for the next month, June 2020. We observe that there are quite a few counties now colored that were not affected yet in May; see the areas encircled by ellipses, but there are also new counties appearing in already affected regions.

All these newly affected counties are counties below the average bar in Fig. 2 and so the pattern was able to predict their destiny. In fact we observed that for 98% of all our patterns the death rate growth was 2-3 times higher than the US average; the other 2% grew at the average pace, none slowed in growth below the US-average. These trends continued in July. This shows that our patterns are highly predictive, and at the same time can also explain the socio-economic conditions for higher-than-average COVID-19 death rate in an easy to understand manner, in the language of the features.

### 3.3 Our Interactive Visual Dashboard

Our dashboard is designed for people with varying levels of visualization literacy to help them navigate and examine the patterns mined with our approach. The dashboard consists of four main panels - geomap, pattern browser, pattern information, and county information - that are linked to each other. Each of these panels are explained in the sections below and also in the caption of Fig. 1. The data input is a standard CSV file with the data matrix.

**Risk Pattern Browser:** As discussed in Section 3.1 patterns are low-D hypercubes, however a collection of patterns can still span a large number of dimensions. This makes it difficult if not impossible to devise an easy to understand visual representation to explain a pattern in its entirety. Thus we choose to represent the collection of patterns as a list of tiles with each tile representing a pattern as shown in the bottom panel in Fig. 1. The patterns are ordered from left to right and top to bottom in descending order of the death rate. This is re-enforced by coloring the tiles from dark to light blue based on the COVID-19 death rate. Only the tiles that pertain to a county selected in the geomap are shaded (more on this below).

Each of these tiles can be clicked which will then trigger updates to the *geomap view* to indicate counties to which this pattern belongs and to the *pattern information* panel to communicate the pattern details (discussed below). Additionally, we change the shape of a selected tile to a circle and give it a yellow border to make it easy for users to locate the selected tile while their focus switches between different elements of the dashboard.

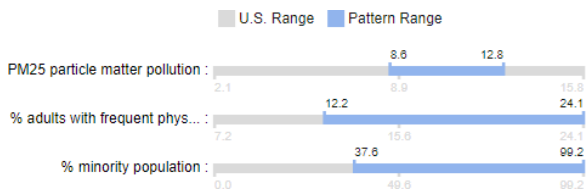

Figure 4: The representation of the ranges of features that define a pattern where the gray bar represents the range of the feature across the US and the blue bar indicates the range for that pattern. For example the third feature '% minority population' has a range of 0 to 99.2 across the US but the range of 37.6 to 99.2 is one of the features of this pattern that drives a higher than average death rate.

**Pattern Information:** This panel communicates the pattern information to the user. A pattern is essentially a set of attributes with specific ranges. Thus the *pattern information* panel reports these ranges to the user while placing them in the context of the global range of the dimension across all data points, in this case all counties. To visualize these ranges, we make use of a bullet chart style visualization that has been shown to be easy-to-understand by a wide audience [15]. An example is shown in Fig. 4. Here each dimension's range is represented as a horizontal bar. The gray portion of the bar indicates the range across all counties in the US of the dimension while the blue portion of the bar represents the range of the dimension that defines this pattern. User can quickly scroll through these bars and study the various ranges that define patterns.

**Geomap View:** This panel consists of an interactive county-level map of the United States (shown in Fig. 1). Each county in the map is assigned a color based on its COVID-19 death rate. We use a continuous color scale ranging from dark blue for high death rates to white for a death rate of zero. Users can click on a county to learn more about the factors leading to its COVID-19 death rate. Clicking on a county will trigger an update to the *risk pattern browser* which highlights the patterns that the county belongs to and grays out the rest. Additionally, the *county information* panel is updated with the top risk features for that county. We also allow the user to zoom and pan the map in order to select smaller counties.

**County Information:** This panel communicates the information of a county selected by the user via the geomap. As shown in the top right corner of Fig. 1 the panel reports the current COVID-19 county death rate as well as the death rate over time. More importantly the panel communicates the top 3 risk factors for the county. Here the feature ranking is computed based on the frequency at which those features appear across all patterns that contain the selected county. We use the same bullet chart-like visualization used for the pattern information to visually represent these features. Here again the gray bars indicate the range of the dimension across all counties in the US while the blue bars are ranges of the features across all counties in the selected county's state. In addition to the ranges shown in the chart we also add markers for the value of the factor in the county and the value for the US average. An example is shown in Fig. 5.

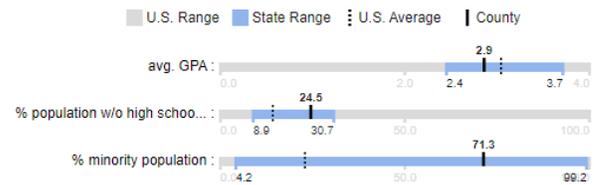

Figure 5: A visual representation of the top 3 feature values of a county in the context of state and US ranges. The gray bar represents the range of the feature across the US and the blue indicates the range of that feature across all counties in the county's state. For example, the county shown here has an 'avg. GPA' of 2.9 (solid black marker) which is slightly lower than the US average (dotted black marker). Additionally the US range for the 'avg. GPA' is 0 to 4 while the range of this feature across all counties in the state is 2.4 to 3.7.

## 4 CONCLUSIONS

We have outlined a methodology that can group socio-economic indicators of public health into 1-3 factor patterns learnt from observational data. The patterns can be used by policy makers and health officials to explain and predict the underlying risk a certain community has with respect to some natural hazard or public health emergency. To give easy access to these patterns we devised an interactive visual dashboard by which the patterns can be explored in the context of the communities' geographical locations. While we have used the early stages of the COVID-19 pandemic to show an application of our methodology, we believe that its application is far broader, which is being explored in ongoing work.

While we provide temporal context to our data – the COVID-19 death rate over time - users currently cannot "roll back" time to examine the patterns at the selected time frame. This is a fairly easy implementation and we plan to add this feature in the future. In addition, while we have used a small cohort of users to gain feedback during system development, we plan a broader task-based study in the near future to gain further insight into utility and usability,

**ACKNOWLEDGMENTS**

This research was funded by NSF SBIR contract 1926949.

**APPENDIX**

In this case study we follow Bob, a public health analyst, who uses the COVID-19 Risk Explorer to learn more about the susceptibility of local communities to the spread of the COVID-19 virus. It's December 2020 and a lot has happened. He starts up the program and sees the screen below.

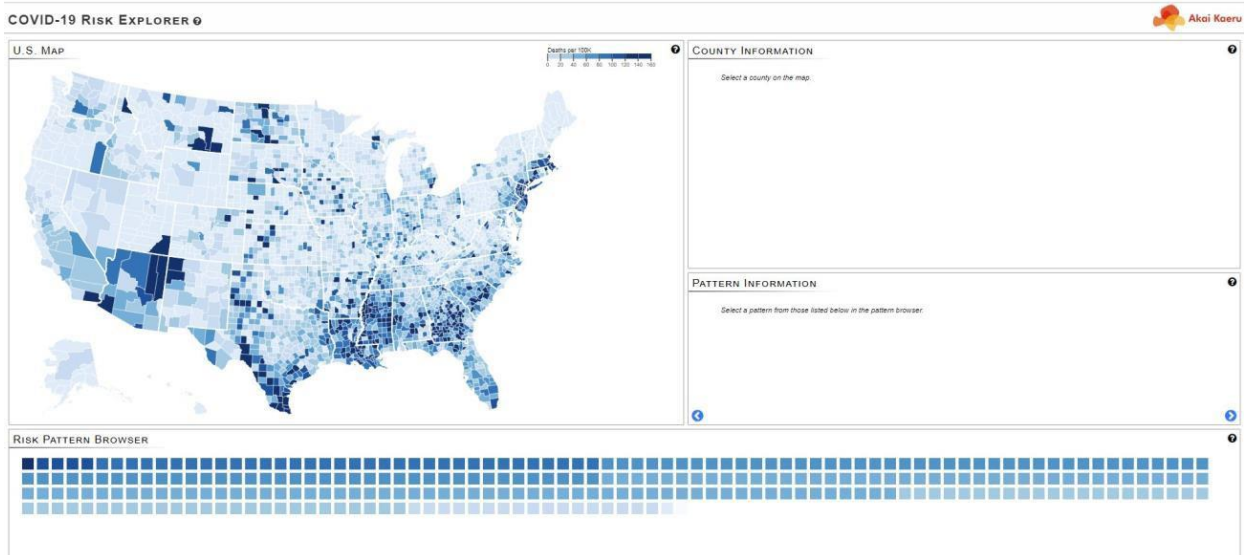

Bob observes that the areas that have seen the highest death rates overall are in Texas, Arizona, Montana, North Dakota, Idaho, the South, Florida, and the Northern East Coast. Now Bob wonders about the timelines. He selects a darkly colored county (a country with high death rates) in **Connecticut – Hartford County**.

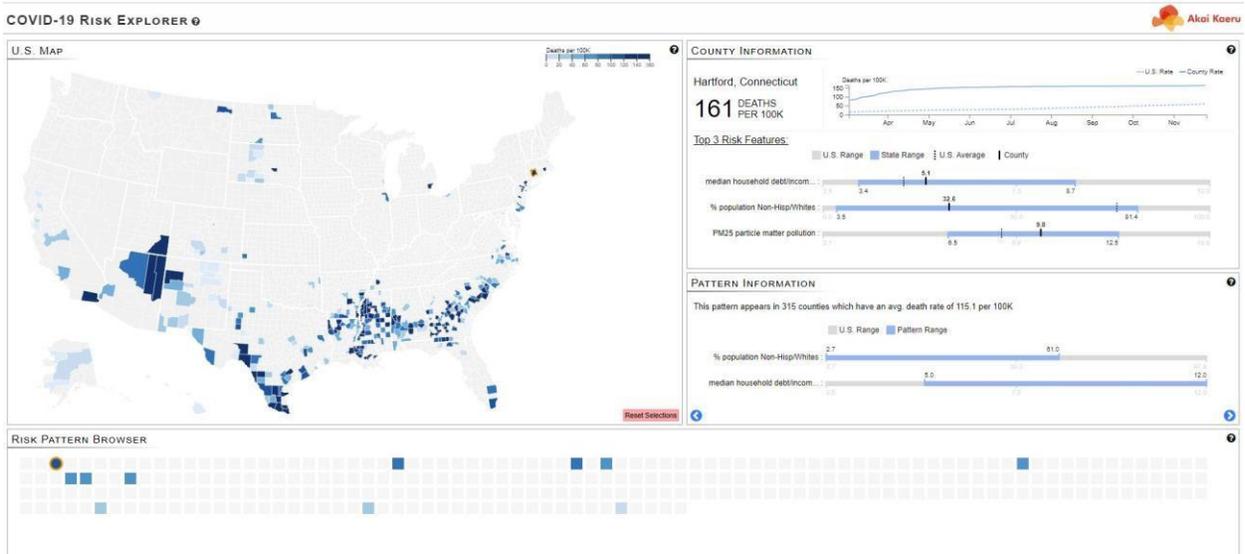

When the screen transitions, Bob observes from the line chart showing the death rate over time in the "County Information" panel that this county exceeded the US average death rate very early in 2020 and quite rapidly so, but then remained nearly flat starting April. Apparently this county responded well and took good precautions to stem the spread.

Bob looks at the "Pattern Information" panel to examine the risk factors of the most dominant pattern Hartford, CT participates in. The "Risk Pattern Browser" indicates that this is the 3[rd] most dominant pattern in the database. It is a 2-feature pattern that indicates that in Hartford, CT the percentage of non-Hispanics/Whites is at the low end of the US overall range and the ratio of median household debt/income is on the high end of the US overall range. Looking at the "Top 3 Risk Factors" of Hartford in the "County Information Panel" Bob learns that these two risk factors are actually the top 3 for Hartford in addition to PM25 particle matter pollution and all its values compare unfavorably to the US average, and while they are not at the extreme ends for the State of Connecticut they tend to be in the unfavorable value ranges.

Next Bob clicks on an equally dark colored county in Southern Texas, **Cameron County, TX** and the screen transitions to what is shown below.

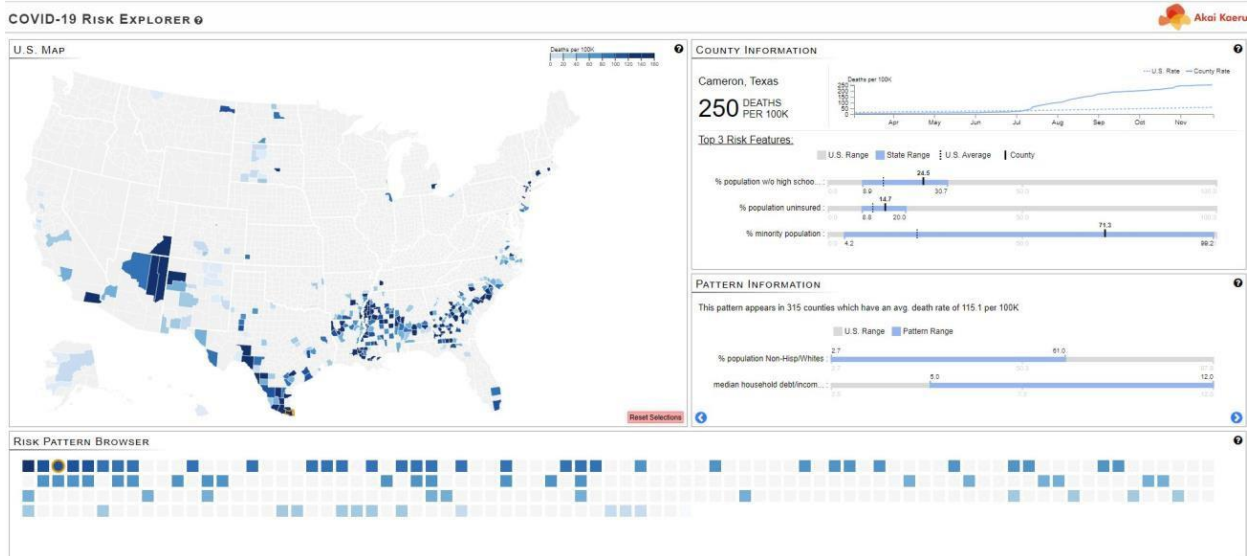

Bob observes that for that county the death rate started to climb much later, in July 2020, and that it is still climbing now in December, albeit at a shallower slope. While Cameron TX shares the 3rd risk pattern with Hartford CT, it has many more risk patterns than Hartford CT, as can be seen by the many filled squares in the "Risk Pattern Browser". It means that its conditions for high death rates are more urgent than for Hartford CT. In fact, its top 3 risk factors are not those of the 3rd risk pattern. For all of these Cameron TX fares unfavorably both within the value range found in Texas and with respect to the US average.

Next, Bob clicks on the first risk pattern in the "Risk Pattern Browser" and sees the screen below.

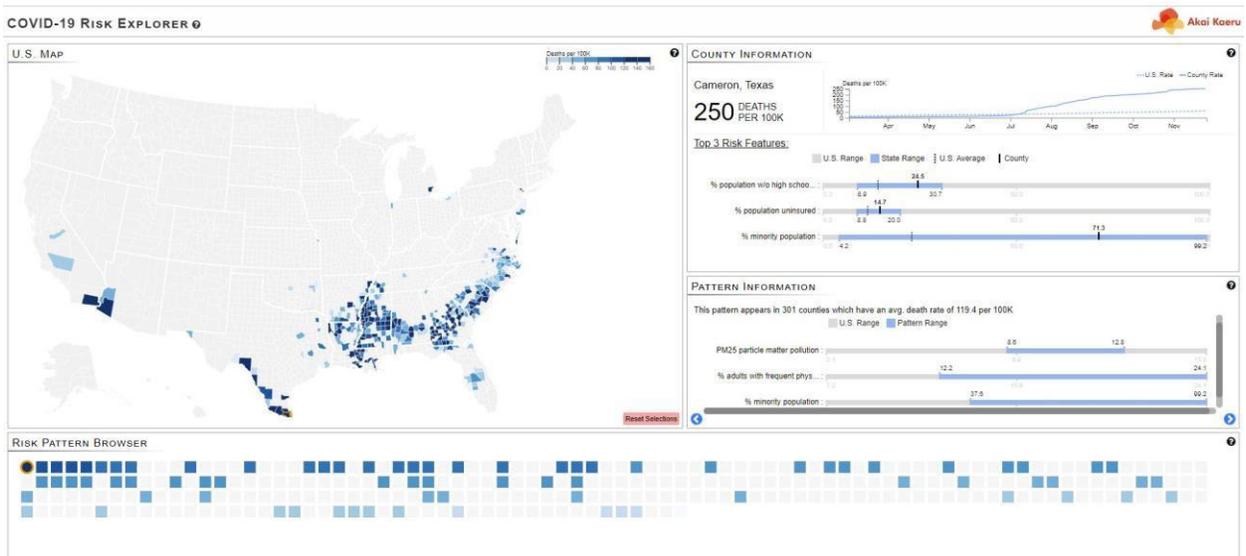

This pattern is a 3-feature pattern with high and unfavorable value ranges in all three of these features. Only the feature "% minority population" appears in the top 3 list for Cameron County, TX and upon further investigation Bob finds that the other two risk features, "% population without high school degree" and "% uninsured", are in risk pattern #4 (see picture next page).

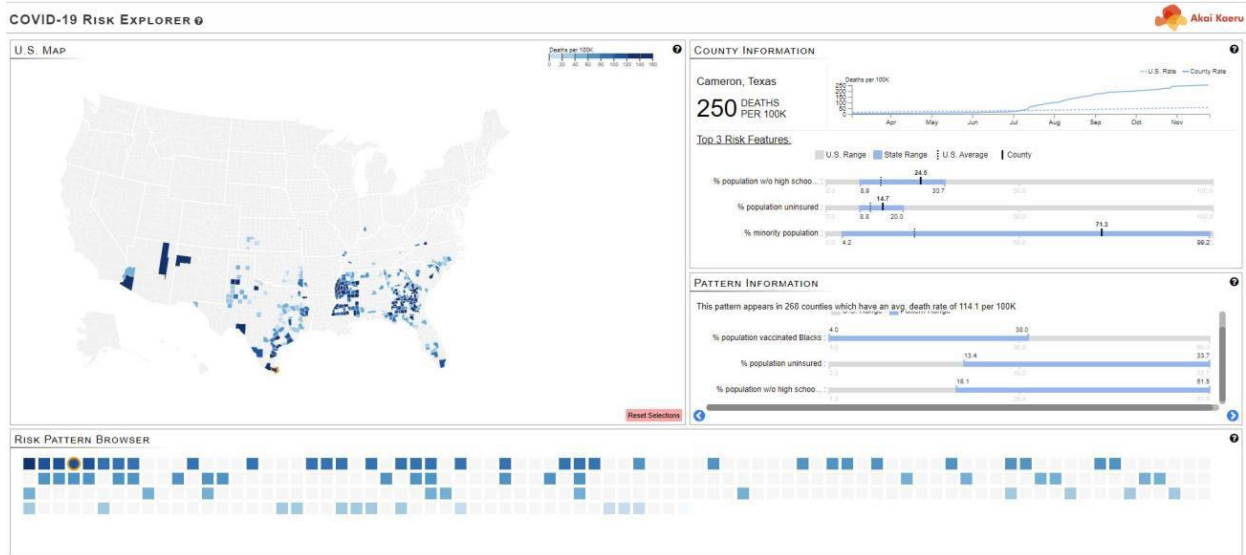

Bob now wants to investigate whether Cameron County, TX could have learned its fate from other counties with similar risk factors but which had experienced high COVID-19 death rates earlier. He looks for counties that share some or all of its top 3 risk features. So he examines counties with risk pattern #1 and risk pattern #4.

He starts with risk pattern #1, clicks a few a counties on the map and eventually learns about **Passaic County, NJ**.

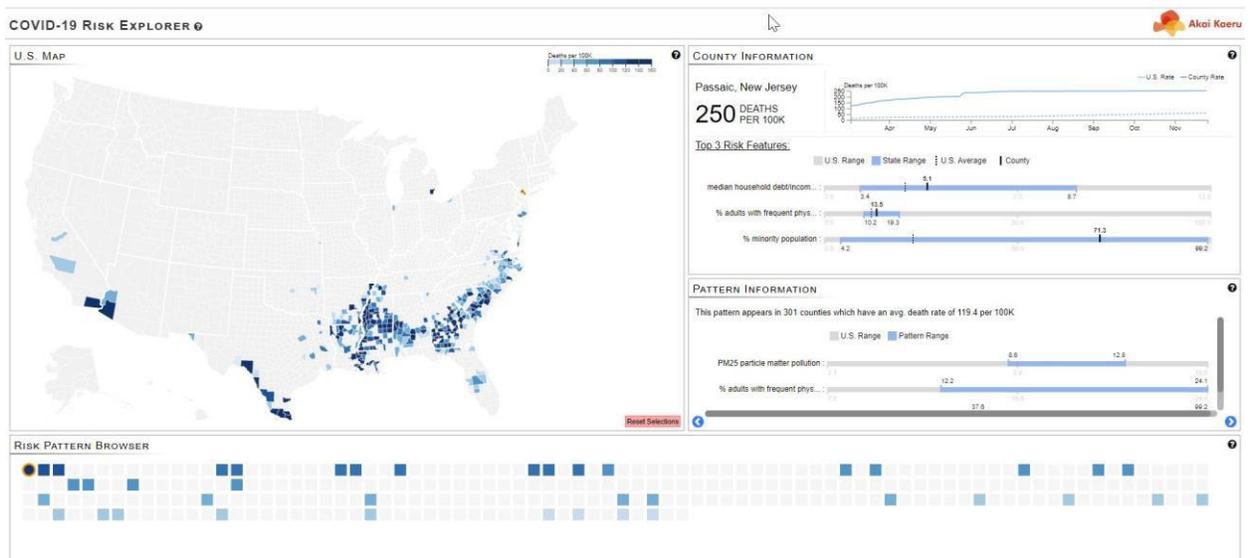

From the timeline he sees that Passaic County, NJ started its death rate at the earliest time and has a similar "% minority population". But this is only one out of the three top 3 risk factors of Cameron County, TX so Bob needs to search more to complete the case. He turns to risk pattern #4 and after some search finds **McKinley, NM** (see next page).

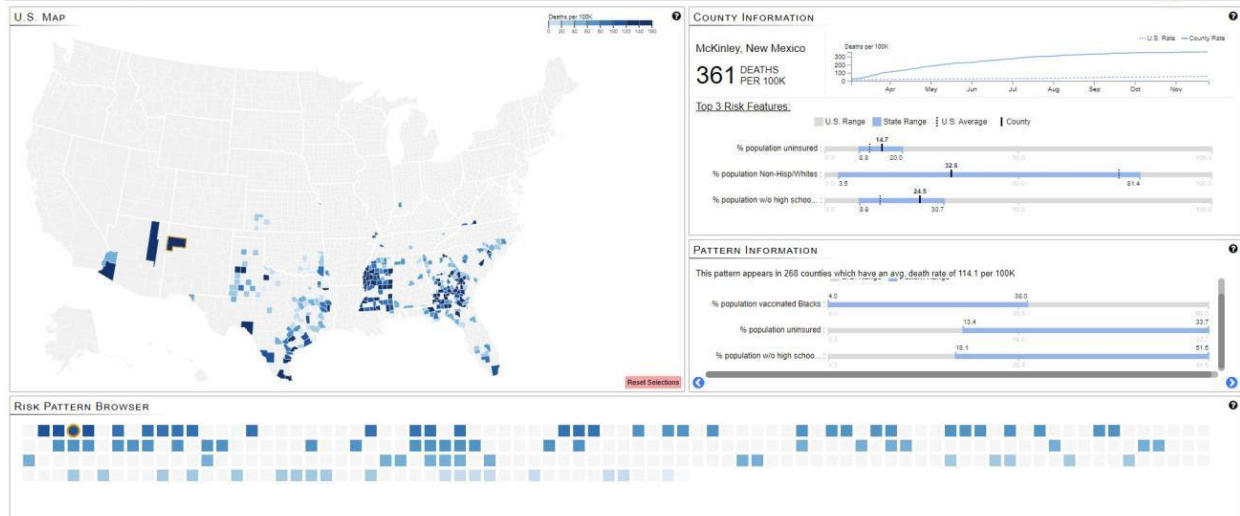

McKinley, NM started its death rate climb later than Passaic, NJ but still 4 months earlier than Cameron, TX. Its top 3 risk factors contain the two other risk factors of Cameron County, TX. So had Cameron County, TX looked at the fate of McKinley County, NM and Passaic County, NJ it could have learned from them and adopt the precautionary measures they took.

There are many more case studies like this one, where late risers could have learnt from early risers about their fate and prepared better. As seen from the early risers' timelines, all of them were able to get the spread under control. Late risers could have taken similar measures and potentially save lives.

As a conclusion, our results show that pattern analysis is a powerful tool for public health risk management and that a dashboard like our Risk Explorer makes it easy to recognize risks and see how outbreaks and disaster in some communities can quickly inform other communities that fit a similar vulnerability pattern to prevent further loss.